\documentclass[aps,showpacs,psfig,floatfix]{revtex4}
\usepackage{graphicx}
\usepackage{epsfig}
\begin{document}

\title{Brownian Motion Model of Quantization Ambiguity and Universality in
Chaotic Systems}
\author{L. Kaplan}
\affiliation{Department of Physics, Tulane University, New Orleans, LA 70118
USA}

\begin{abstract}
 We examine spectral equilibration of quantum chaotic spectra to universal
statistics, in the context of the Brownian motion model.  Two competing time
scales, proportional and inversely proportional to the classical relaxation
time, jointly govern the equilibration process.  Multiplicity of quantum
systems having the same semiclassical limit is not sufficient to obtain
equilibration of any spectral modes in two-dimensional systems, while in
three-dimensional systems equilibration for some spectral modes is possible if
the classical relaxation rate is slow.  Connections are made with upper bounds
on semiclassical accuracy and with fidelity decay in the presence of a weak
perturbation.
\end{abstract}

\pacs{05.45.Mt, 03.65.Sq}

\maketitle

\section{Introduction}
\label{secintro}

 Random Matrix Theory (RMT) was introduced by Wigner in the 1950s as a model
for describing the universal spectral behavior of complex many-body systems
such as compound nuclei.  It was conjectured that a typical spectrum of
a complex system displays statistical properties similar to those of
a Hamiltonian chosen at random from a basis-independent ensemble of random
matrices in the same symmetry class.  Subsequently, Dyson showed that
the statistical predictions of RMT could be reproduced within a Brownian
motion model, where Brownian motion of the energy levels results from a
stochastic perturbation of some initial, possibly nonrandom,
Hamiltonian~\cite{dyson}.  Bohigas, Giannoni, and Schmit argued that the same
universal behavior of the spectrum occurs generically even in single-particle
systems, as long as the underlying classical dynamics is chaotic~\cite{bgs}.

 Wilkinson used the Brownian motion model to explain the connection between
quantum chaotic systems and RMT spectra, using the ``quantization ambiguity"
(i.e, multiplicity of quantum systems having the same semiclassical limit) as
the source of perturbation causing Brownian motion of the energy levels and
eventual equilibration to RMT spectral statistics~\cite{wilk}.  More recently,
the Brownian motion approach has been extended to study the evolution of
eigenstates~\cite{wilkwalk} and to relate the smooth and oscillatory parts of
the spectral correlation function in diffusive and ballistic chaotic
systems~\cite{mehlwilk}.

 It well known that not all spectral modes of a typical quantum chaotic
spectrum obey universal statistics.  It is therefore of interest in the context
of the Brownian motion model to investigate which modes are equilibrated in a
given dynamical system, for a given perturbation strength (such as that arising
from the quantization ambiguity), as a function of degree of chaos in the
system, and as a function of effective $\hbar$.  Alternatively, one may ask
about the size of perturbation necessary to achieve equilibration to universal
statistics for a given spectral mode; this question is related to work by
Zirnbauer on energy level correlations in a quantum ensemble of classically
identical maps~\cite{zirnbauer}, as well as to the method developed by Gornyi
and Mirlin for studying wave function correlations in ballistic systems by
adding weak smooth disorder~\cite{gornyi}.

 In the present work, we show that the Brownian motion model can be used
successfully to predict the number of equilibrated spectral modes as a function
of perturbation strength, chaoticity of the classical system dynamics, system
dimension, and effective $\hbar$.  The equilibration process is governed by two
competing time scales, allowing the number of modes equilibrated to RMT either
to increase or to decrease as the original system becomes more chaotic;
equilibration is reduced for systems with very short or very long classical
relaxation times and is maximized for systems with intermediate Lyapunov
exponent.  We show that in two-dimensional quantum systems no equilibration to
universal behavior for any spectral modes can arise from the quantization
ambiguity in the $\hbar \to 0$ limit; in three dimensions some modes may be
equilibrated depending on the classical relaxation time of the underlying
classical dynamics; in four dimensions and higher, equilibration to
universality will occur for at least some modes, but maximal equilibration
requires the classical relaxation time to be much longer than the typical
ballistic time scale of the dynamics. 

 The paper is organized as follows.  In Sec.~\ref{secvar}, we review
semiclassical estimates for matrix element variance, which are required as
input for the Brownian motion model.  In Sec.~\ref{secmaps}, we discuss several
example systems used for comparison between theory and numerics, and in
Sec.~\ref{secbrownian} we review the Brownian motion model itself.  Two
competing conditions for equilibration of spectral modes are obtained in
Sec.~\ref{secequil} and applied to equilibration of the nearest level spacing
distribution in Sec.~\ref{secspac}.  The implications for the effect of
quantization ambiguity on spectral universality are developed in
Sec.~\ref{secambig}.  Finally, in Sec.~\ref{secsemi} we demonstrate the
relationship between results obtained in the Brownian motion model and recent
findings on semiclassical accuracy~\cite{sckaplan} and decay of quantum
fidelity~\cite{prosen}.

\section{Semiclassical Calculation of Matrix Element Variance}
\label{secvar}

 This discussion follows~\cite{wilk,wilkwalk,eckh}.  Consider the matrix
elements $B_{nn}=\langle \Psi_n |\hat B|\Psi_n \rangle$ of an operator $\hat
B$, given the eigenstates $|\Psi_n \rangle$ of Hamiltonian $\hat H$.  It is
assumed that the operator $\hat B$ is ``independent" of $\hat H$ and has a
well-defined classical limit $B(q,p)$.  We also assume that the phase-space
average of the corresponding classical observable vanishes, so that the matrix
elements fluctuate around zero, $\overline{B_{nn}} =0$.  The variance of the
matrix elements at energy $E$ is
\begin{equation}
\sigma^2_B(E) = \overline{B_{nn}^2} = \nu(E)^{-1} \overline{\sum_n B_{nn}^2
\delta_\eta (E-E_n)} \,,
\end{equation}
where $\delta_\eta(E)=(2 \pi \eta^2)^{-1/2} e^{-E^2/2\eta^2}$,
$\nu(E)=\overline{\sum_n \delta_\eta(E-E_n)}$ is the density of states, and
$\overline{\cdots}$ denotes averaging over a ``suitable" ensemble.  One may
approximate
\begin{equation}
\sigma^2_B(E) \approx {2 \sqrt{\pi} \,\epsilon \over \nu(E)}
\overline{\left [ \sum_n B_{nn} \delta_\epsilon(E-E_n) \right ]^2} \,,
\end{equation}
where $\epsilon = \sqrt{2}\, \eta$ is of the order of but smaller than the mean
level spacing $\Delta=1/\nu(E)$.  In the semiclassical limit, the expression in
square brackets is approximated by a trace formula
\begin{equation}
\left [ \cdots \right ] \approx {1 \over 2 \pi\hbar}
{\rm Im} \sum_{p,r} T_{p} \,
B_p \, D_{p,r} \,  e^{ i S_{p,r}/\hbar - i \mu_{p,r} \pi/2} \,
e^{-\epsilon^2 r^2 T_p^2 / 2 \hbar^2} \,.
\end{equation}
Here the sum is over all primitive periodic orbits $p$ at energy $E$ and their
repetitions $r$, $T_p$ is the primitive orbit period, $B_p$ is the classical
average of the observable $B$ over the orbit, $D_{p,r}= |{\rm
det}(M_{p,r}-1)|^{-1/2}$ is the square root of a classical focusing factor,
$M_{p,r}$ is the monodromy matrix of the orbit, $S_{p,r}$ is the action, and
$\mu_{p,r}$ is the Maslov index.  Using $\overline{({\rm Im} f)^2} = {1 \over
2} {\rm Re} \overline{f^\ast f}$ and neglecting repetitions, one obtains the
following estimate for the variance,
\begin{equation}
\sigma^2_B(E) \approx {2 \over \beta} {2 \sqrt{\pi} \, \epsilon \over \nu(E)}
(2 \pi \hbar)^{-2}  \sum_p T_p^2 |B_p|^2 D_p^2 \, e^{-\epsilon^2 T_p^2 /
\hbar^2} \,,
\end{equation}
where as usual $\beta=1$, $2$ characterizes the presence or absence of time
reversal symmetry, respectively.  Now, assuming $B_p$ to be real, and averaging
over many periodic orbits of period close to $T_p$,
\begin{equation}
B_p^2 = T_p^{-2} \int_0^{T_p} dt_1 \int_0^{T_p} dt_2 \,
B(t_1) B(t_2) \approx {1 \over T_p} \int_{-T_p/2}^{T_p/2} dt \,
\langle B(0) B(t) \rangle\,,
\end{equation}
where $\langle B(0) B(t) \rangle$ is the classical average of
$B(q,p)B(q(t),p(t))$ over the energy hypersurface at energy $E$.  Making use of
the classical sum rule
\begin{equation}
\sum_p T_p \, f(T_p) \, D_p^2 \, e^{-\epsilon^2 T_p^2 /\hbar^2}
\approx  2\int_0^\infty dT \, f(T) \, e^{-\epsilon^2 T^2/\hbar^2} \,,
\end{equation}
one finally obtains
\begin{eqnarray}
\sigma^2_B(E) &\approx & {2 \over \beta} {4 \sqrt{\pi} \, \epsilon \over \nu(E)}
(2 \pi \hbar)^{-2} \, \int_0^\infty dT \int_{-T/2}^{T/2} dt \,
\langle B(t)B(0) \rangle e^{-\epsilon^2T^2/\hbar^2} \nonumber \\
&=& {2 \over \beta} { 1 \over \pi  \nu(E)  \hbar} \int_0^\infty dt \,
\langle B(t)B(0) \rangle f_\epsilon(t) \,,
\end{eqnarray}
with $f_\epsilon(t)=1-{\rm erf}(2\epsilon t/\hbar)$.  This may be
approximated~\cite{artuso} as
\begin{equation}
\label{sigmaint}
\sigma^2_B(E) \approx {2 \over \beta} { \langle B^2 \rangle \over \pi \nu(E)
\hbar} \int_0^\infty dt \, P(t) f_\epsilon(t) \,,
\end{equation}
where
\begin{equation}
\label{pt}
P(t)={\langle \, B(t) \, B (0)\,\rangle \over \langle \, B^2 \, \rangle}
\end{equation}
is a classical correlation function.  For a chaotic system, $P(t)$ typically
decays as a sum of exponentials~\cite{artuso}.  The long-time behavior is then
governed by the smallest exponent $\lambda$, $P(t) \approx b \, e^{-\lambda
t}$, where $b$ is a classical constant of order unity.  If the long-time
behavior dominates the integral, one obtains in the limit $\epsilon \ll \lambda
\hbar$
\begin{equation}
\label{longdom}
\sigma^2_B(E) \approx {2 \over \beta} { b\, \langle B^2 \rangle \over \pi
\nu(E) \hbar  \lambda }
= {4 b\,\langle B^2 \rangle \over \beta  \lambda T_H}
= {4 b\,\langle B^2 \rangle\, T_{\rm decay} \over \beta T_H}
\,,
\end{equation}
where $T_H=2\pi\hbar/\Delta=2\pi\hbar \nu(E)$ is the Heisenberg time, at which
individual eigenstates are resolved, and $T_{\rm decay}=\lambda^{-1}$ is a
classical correlation decay time.  Since $\nu(E) \propto \hbar^{-d}$, the
variance decays as
\begin{equation}
\label{varhlam}
\sigma^2_B(E) \propto \hbar^{d-1}
\lambda^{-1}
\end{equation}
for small $\hbar$.  This is consistent with Shnirelman's conjecture concerning
the ergodicity of chaotic eigenstates.  The result is independent of the
smoothing function $\delta_\epsilon(E)$.  In the limit of small $\lambda$, the
decay time $T_{\rm decay}=\lambda^{-1}$ becomes longer than the Heisenberg time
$T_H = 2 \pi  \hbar  \nu(E)$, and $\lambda^{-1}$ in Eq.~(\ref{varhlam}) should
be replaced with $2 \pi \hbar  \nu(E)$.  Then $\sigma^2_B(E) \propto \hbar^0$,
as expected for a quantum regular system.  We can interpolate between the
extreme chaotic and effectively regular regimes by taking $T_{\rm decay}$
intermediate the period of the shortest orbit and the Heisenberg time.

\section{Classically Chaotic Maps}
\label{secmaps}

 As an example, consider the class of maps
\begin{eqnarray}
p_{k+1} &=& p_k-V'(q_{k}) \nonumber
 \\
q_{k+1}&=&q_k+T'(p_{k+1}) 
\,,
\end{eqnarray}
on the torus $[-\pi,\pi) \times [-\pi,\pi)$.  Any such map may be thought of as
arising from the periodically kicked Hamiltonian~\cite{kick,kick2}
\begin{equation}
H(p,q,t) = { 1 \over T_{\rm kick}}
T(p) + V(q) \sum_{j=-\infty}^\infty \delta(t-j \, T_{\rm kick}) \,.
\end{equation}
The choice $T(p)=v \, p^2/2+K_2 \, v \sin p$ and $V(q)=-v \, q^2 /2 - K_1 \, v
\, \sin q$ corresponds to a perturbed sawtooth map.  Here
\begin{equation}
\label{sawtooth}
\begin{array}{lll}
p_{k+1} &=& p_k+v\, q_k+ K_1 \, v \, \cos q_k  \nonumber \\[5pt]
q_{k+1} &=& (1 + v^2) q_k + v \, p_k+K_1 \, v^2 \, \cos q_k 
+ K_2 \, v \, \cos p_{k+1}
\end{array}
\end{equation}
and linearized motion is governed by the monodromy matrix
\begin{equation}
M = \left [ \begin{array}{cc} 
1 & v\, (1-  K_1 \sin q_k) \\ v\, (1-K_2 \sin p_{k+1}) &
1+v^2\, (1-K_1 \sin q_k)(1-K_2 \sin p_{k+1}) 
\end{array} \right ] \,.
\end{equation}
In this case, strict hyperbolicity is guaranteed for $v \ne 0$, as long as
$|K_1|<1$ and $|K_2|<1$.  The parameters $K_1$ and $K_2$ introduce nonlinearity
and symmetry breaking into the dynamics.  The classical Lyapunov exponent and
correlation decay time $T_{\rm decay}$ may easily be controlled by adjusting
the parameter $v$.

\begin{figure}[ht]
\centerline{
\psfig{file=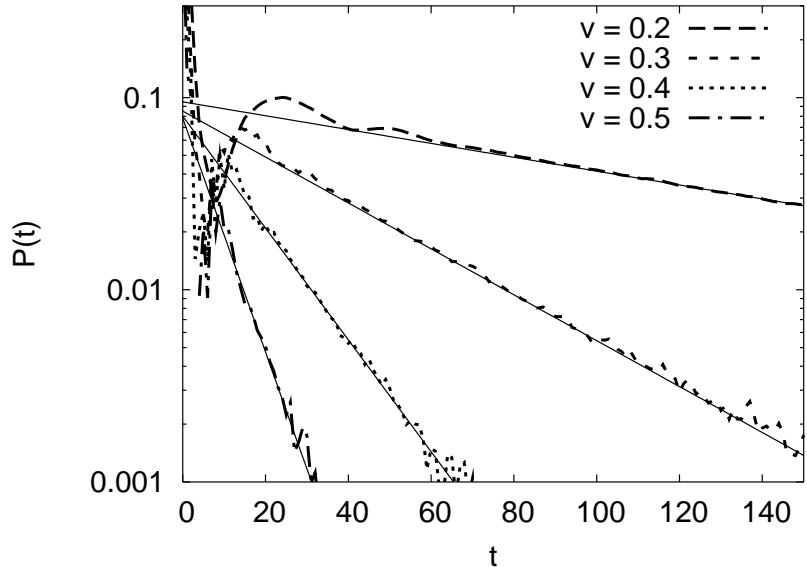,angle=270,width=4.5in}}
\vskip 0.2in
\caption{
The classical autocorrelation function $P(t)$ of Eq.~(\ref{pt}) is plotted for
the perturbed sawtooth map of Eq.~(\ref{sawtooth}), with $K_1=0.3$, $K_2=-0.2$,
and several values of $v$.  $P(t)$ is computed for each of the six observables
$B(q,p)=\cos (2 m \pi q), \, \sin(2 m \pi q)$, where $m = 1$, $2$, $3$, and
then averaged to obtain the behavior for a typical observable.  The thin
solid lines represent best fits to the long-time exponential behavior $P(t)=b
e^{-\lambda_v t}$, with $\lambda_{0.2}=0.0083$, $\lambda_{0.3}=0.0275$,
$\lambda_{0.4}=0.067$, and $\lambda_{0.5}=0.14$.  } \label{fig_clas}
\end{figure}

 In Fig.~\ref{fig_clas}, the classical autocorrelation function $P(t)$ for the
perturbed sawtooth map is shown for several values of the $v$ parameter.
Clearly, the long-time decay exponent $\lambda$ decreases with decreasing $v$,
and empirically we find $\lambda_v \sim v^{3.1}$ for $v \ll 1$.  Note that the
correlation decay rate $\lambda$ behaves very differently from the Lyapunov
exponent, which scales linearly with $v$.

 When a discrete-time map is quantized on an $N-$dimensional Hilbert space, we
have $\hbar=2 \pi/N$, $\nu(E)=(N/2 \pi)^2 \, T_{\rm kick}$, and $T_H=N \,
T_{\rm kick}$.  Then Eq.~(\ref{sigmaint}) becomes
\begin{equation}
\label{sigmaintmap}
\sigma_B^2(E) \approx {2 \over \beta} {2 \, \langle B^2 \rangle \over N}
\left [ {1 \over 2} + \sum_{k=1}^{\infty} P(k) f_\epsilon(k)\right ] \,,
\end{equation}
and Eq.~(\ref{longdom}) takes the form
\begin{equation}
\sigma_B^2(E) \approx {2 \over \beta} {2 \, b\langle \, B^2 \rangle \over
\lambda \, N \, T_{\rm kick}} \,.
\end{equation}

 We may also consider a system with power-law decay of classical correlations,
e.g.,
\begin{eqnarray}
\label{clpower}
T(p) &=& {1 \over 2}p^2+\cos (2 \pi p) \nonumber \\
V(q) &=& \left \{ \begin{array}{ll} -{1 \over 2} \left [ q^2 - \pi^2/4 \right ]
\;\;\; & {\rm for} \; |q| <\pi/2 \\[5pt] 0 \;\;\; & {\rm otherwise} \end{array} \right .
\,.
\end{eqnarray}
Here, bouncing-ball modes for $|q|< \pi/2$ in the vicinity of $p=0$ dominate
the long-time classical correlation decay, resulting in $P(k) \sim b \,
k^{-\gamma}$, with $\gamma=1/3$.  The sum over time steps $k$ in
Eq.~(\ref{sigmaintmap}) is effectively truncated on the scale of the Heisenberg
time $N$ by the smoothing function $f_\epsilon(k)$.  Then $\sum_{k=1}^{N/2}
P(k)f_\epsilon(k) \sim N^{2/3}$ and thus
\begin{equation}
\sigma_B^2(E) \sim {\langle B^2 \rangle \over N^{1/3}} \,.
\end{equation}

\section{Brownian Motion Model for Chaotic Energy Levels}
\label{secbrownian}

 The statistical properties of the energy levels $E_n$ can be obtained using a
Brownian motion model~\cite{dyson}.  The discussion in this section follows
\cite{wilk,wilkwalk,mehlwilk}.  In this model, the matrix elements of the
Hamiltonian undergo a diffusive evolution as a function of a fictitious time
variable $\tau$.  We denote the infinitesimal change of $\hat H$ by $\delta
\hat H$.  The off-diagonal elements of $\delta \hat H$ are assumed to satisfy
\begin{equation}
\overline{\delta H_{mn}}=0 \,, \;\;\;\; \overline{\delta H_{mn}
\delta H_{m'n'}^\ast} = \delta \tau \, C_{mn}^{\rm off} \delta_{nn'}
\delta_{mm'}\,,
\end{equation}
while the diagonal elements are taken to obey
\begin{equation}
\overline{\delta H_{nn}}=0 \,, \;\;\;\; \overline{\delta H_{mm}
\delta H_{nn}} = { 2 \over \beta} \, \delta \tau \, C_{mn}^{\rm diag} \,.
\end{equation}
Within RMT, $C_{mn}^{\rm diag} = \delta_{mn}$ and $C_{mn}^{\rm off} =1$.
Non-universal deviations from RMT are encoded in $C_{mn}^{\rm
diag}=C_{m-n}^{\rm diag}$ and $C_{mn}^{\rm off}=C_{m-n}^{\rm
off}$~\cite{mehlwilk}.  The corresponding motion of the energy levels $E_0$ ,
...  , $E_{N-1}$ may be analyzed as follows.  Using second order perturbation
theory, one obtains
\begin{equation}
\delta E_n = \sum_{m \ne n} {|\delta H_{mn}|^2 \over E_n -E_m}
+\delta H_{nn}
\end{equation}
for the energy level shifts $\delta E_n$.  Thus
\begin{equation}
\label{deltaen}
\overline{\delta E_n} = \delta \tau \sum_{m \ne n}
{C_{m-n}^{\rm off} \over E_n -E_m}
\end{equation}
and
\begin{equation}
\label{deltaemn}
\overline{\delta E_m \delta E_n} = {2 \over \beta} \, \delta \tau \,
C^{\rm diag}_{m-n} \,.
\end{equation}

 It is convenient to express Eqs.~(\ref{deltaen}) and (\ref{deltaemn}) in terms
of the Fourier modes of $\Delta E_n \equiv E_n - n \Delta$.  We enforce
periodic boundary conditions $\Delta E_N = \Delta E_0$ and define
\begin{equation}
a_k = {1 \over N} \sum_{n=0}^{N-1} \Delta E_n e^{-2 \pi i kn/N}
\end{equation}
so that 
\begin{equation}
\Delta E_n = \sum_{k=-N/2}^{N/2-1} a_k \, e^{2 \pi i k n /N} =
\sum_{k=0}^{N/2-1} a_k \, e^{2 \pi i k n /N} + {\rm c.c.}
\end{equation}
Since the $\Delta E_n$ are real, $a_k=a^\ast_{-k}$.  From Eq.~(\ref{deltaemn})
we have, using $\delta \Delta E_n = \delta E_n$,
\begin{equation}
\label{dakdap}
\overline{\delta a_k \delta a_p^\ast} =
\overline{\delta a_k^\ast \delta a_p}={2 \over \beta} \,
\delta \tau \,  I_k \, \delta_{kp}  \,,
\end{equation}
where $I_k=N^{-1} \sum_n C_n^{\rm diag} e^{2 \pi i kn/N}$, and
\begin{equation}
\overline{\delta a_k \delta a_p} = \overline{\delta a_k^\ast \delta a_p^\ast}
=0
\end{equation}
for $k,p \in (0,N/2)$.  In general, the expectation value of $\delta a_k$ is a
complicated function of all $a_p$, which may be expanded as
\begin{equation}
\label{daexpand}
\overline{\delta a_k} = \delta \tau
\left [ 
A_1(k) a_k + \sum_q A_2(k,q) a_{k-q}a_q +\sum_{q,p} A_3(k,q,p) a_q a_p
a_{k-q-p} + \cdots \right ] \,.
\end{equation}
The coefficients $A_j$ are obtained as follows (assuming that $C_{m-n}^{\rm
off}=1$, see however Ref.~\cite{chalker}).  Expanding the denominator $E_n-E_m$
in Eq.~(\ref{deltaen}) around the mean $(n-m)\Delta$,
\begin{equation}
\overline{\delta E_n} \approx -\delta \tau
\sum_{\ell \ne 0} { 1\over \ell \Delta}
\left [ 
1- {1 \over \ell \Delta}(\Delta E_{n+\ell}-\Delta E_n)
+ {1 \over 2 \ell^2 \Delta^2}(\Delta E_{n+\ell}-\Delta E_n)^2 +\cdots
\right ] \,.
\label{deltaenexp}
\end{equation}
Again using $\delta E_n = \delta \Delta E_n$ and the identity $ {1 \over N}
\sum_n e^{-2 \pi i kn/N} =\sum_j\delta_{k,j N} $, the first term in
Eq.~(\ref{deltaenexp}) implies
\begin{equation}
\overline{\delta a_k} = \delta \tau a_k \sum_{\ell \ne 0}
\left [ 
{1 \over \ell^2\Delta^2} \left ( e^{2\pi ik \ell /N} -1 \right ) + \cdots
\right ] \,.
\end{equation}

 In the limit of large $N$ one has to first order in $|k|/N$ (extending the
range of summation over $\ell$ from $-\infty$ to $\infty$)
\begin{equation}
\sum_{\ell \ne 0}
{1 \over \ell^2\Delta^2} \left ( e^{2\pi ik \ell /N} -1 \right )
\approx - {2 \pi^2 |k| \over N \Delta^2} \,,
\end{equation}
and thus
\begin{equation}
A_1(k)= - {2 \pi^2|k| \over N \Delta^2} \,.
\end{equation}
To lowest order in Eq.~(\ref{daexpand}), the equilibrium distribution of the
$a_k$ (corresponding to large fictitious time $\tau$) factorizes into a product
of Gaussians, where the variance is given by 
\begin{equation}
\label{ak2}
\overline{|a_k|^2} = {2 \over \beta} {N \Delta^2 I_k \over
4 \pi^2 k} \,,
\end{equation}
while higher-order cumulants are zero.  The approximation of Eq.~(\ref{ak2}) is
appropriate for $k \ll N$.  In this case, the $a_k$ for different $k$ are
uncorrelated.  In RMT,  $I_k=N^{-1}$, and
\begin{equation}
\label{ak2rmt}
\overline{|a_k|^2} = {2 \over \beta} {\Delta^2 \over 4 \pi^2 k}
\end{equation}
for $k \ll N$.

 In order to determine the fluctuations of $a_k$ for $k \sim N$, higher-order
terms in Eq.~(\ref{daexpand}) must be taken into account.  This may be done
perturbatively, resulting in corrections to the variance and possibly non-zero
higher-order cumulants.  Moreover, $a_k$ for larger values of $k$ may be
correlated.

 In the above derivation, we have started with initial conditions $a_k=0$ at
fictitious time $\tau=0$, i.e. we have assumed the initial unperturbed
Hamiltonian has a ``picket fence" spectrum, $E_n = n \Delta$.  A typical
chaotic Hamiltonian, however, will not correspond naturally to a small
perturbation of such a ``picket fence" Hamiltonian.  Therefore, in practice it
is more useful to observe spectral equilibration by comparing coefficients
$a_k$ of a given Hamiltonian $\hat H$ with the coefficients $a_k'$ of a
perturbed Hamiltonian $\hat H+\hat B$.  Full equilibration implies that the
spectra of $\hat H$ and $\hat H +\hat B$ become independent of one another,
though drawn from the same random matrix ensemble.  Then
\begin{equation}
\label{akdiffrmt}
\overline{|a_k-a_k'|^2} = {2 \over \beta} {\Delta^2 \over 2 \pi^2 k}
= {C_\beta \over k}
\end{equation}
for $k \ll N$.

\section{Conditions for Energy Level Equilibration}
\label{secequil}

 We now address more carefully the question originally raised in the seminal
work by Wilkinson~\cite{wilk}.  Specifically, we wish to understand under what
circumstances a class of classically small perturbations $\hat B$ of an initial
Hamiltonian $\hat H$ is sufficient to generate random matrix statistics in the
spectrum, at various energy scales.  A related question is the size in
parameter space of a random chaotic ensemble necessary to average away
system-specific spectral properties and generate universal
statistics~\cite{zirnbauer,gornyi}.

 Two conditions are necessary for equilibration to universal statistics to
occur.  First, if the perturbation is classically small, then equilibration to
RMT may only occur on time scales longer than the decay time of classical
correlations (otherwise, $I_k \ne N^{-1}$ for the corresponding modes $a_k$).
Thus, the modes $a_k$ may be equilibrated to RMT only for $k \gg T_{\rm
decay}/T_{\rm kick}$, independent of the perturbation, and correspondingly the
spectrum may only display universal statistics on energy scales ${\cal E} \ll
\hbar/T_{\rm decay}$.  Secondly, the perturbation must be sufficiently strong
to equilibrate a given mode $a_k$ after the fictitious time $\tau$ during which
the Hamiltonian matrix elements undergo Brownian motion.  In our units, this
fictitious time $\tau$ is simply equal to the variance~\cite{wilk}:
\begin{equation}
\tau = \sigma^2_B \,,
\end{equation}
and is given by Eq.~(\ref{longdom}).  On the other hand, the characteristic
response time of the $k$-th mode is
\begin{equation}
\tau_k = {\hbar \over \pi \nu(E) (k T_{\rm kick})} = {2 \hbar^2 \over
(k T_{\rm kick}) T_H} \,,
\end{equation}
as may easily be seen by comparing Eqs.~(\ref{dakdap}) and (\ref{ak2}) and
then noting that the Heisenberg time $T_H$ is given by
$2\pi\hbar/\Delta=2\pi\hbar \nu(E) =N T_{\rm kick}$.
Equilibration will therefore happen for a given mode $a_k$ when $\tau \gg
\tau_k$, i.e.
\begin{equation}
\label{kcond}
k \gg {\hbar^2 \over \langle B^2 \rangle  T_{\rm decay} T_{\rm kick} } \,,
\end{equation}
and transforming to the energy domain we find equilibration on scales
\begin{equation}
\label{econd}
{\cal E} \ll {\langle B^2 \rangle T_{\rm decay} \over \hbar} \,.
\end{equation}
The condition of Eq.~(\ref{econd}) must be satisfied simultaneously with the
first condition ${\cal E} \ll \hbar /T_{\rm decay}$, i.e. equilibration to
universal statistics occurs on all energy scales $\cal E$ satisfying
\begin{equation}
\label{twoconde}
 {\cal E } \ll {\cal E}_{\rm equil} \sim {\rm min} \left (
{\hbar \over T_{\rm decay}}, {\langle B^2 \rangle T_{\rm decay} \over \hbar}
\right ) \,.
\end{equation}
Note that the two upper limits on the equilibration energy scale have opposite
dependence on the classical correlation scale $T_{\rm decay}$.

 Let us consider the behavior of the system as the size of the perturbation
$\hat B$ is varied, for a given initial Hamiltonian $\hat H$, at a given
classical energy.  We assume $T_{\rm decay} < T_H$, so that the system is in
the quantum chaotic regime.  (i) For very small perturbations, $\langle B^2
\rangle \ll \hbar^2 / (T_{\rm decay} T_H)$, there is no equilibration on any
time scale before the Heisenberg time $T_H$ and consequently no equilibration
on any energy scale larger than a mean level spacing $\Delta$.  (ii) For larger
perturbations, i.e. $\hbar^2 / (T_{\rm decay} T_H) \ll \langle B^2 \rangle  \ll
\hbar^2/T_{\rm decay}^2$, all modes $k \gg k_{\rm min} \sim \hbar^2 / (T_{\rm
kick} T_{\rm decay} \langle B^2 \rangle)$ equilibrate to their RMT values.  In
the energy domain, the corresponding condition is ${\cal E} \ll {\cal E}_{\rm
equil} \sim \langle B^2 \rangle T_{\rm decay} / \hbar$.  (iii) Finally, for the
largest (still classically small) perturbations $\hbar^2 / T_{\rm decay}^2 \ll
\langle B^2 \rangle \ll E^2$, all time scales beyond $T_{\rm decay}$ and all
energy scales below ${\cal E}_{\rm equil} \sim \hbar/T_{\rm decay}$ equilibrate
to universal statistics.

 What happens if we instead fix the size of perturbation $\hat B$, and consider
a variety of classical systems, with different classical time scales $T_{\rm
decay}$?  (i) For the most strongly chaotic systems, $T_{\rm decay} \ll \hbar
{\langle B^2 \rangle }^{-1/2}$, and equilibration is limited by the condition
of Eq.~(\ref{econd}).  (ii) As the classical system becomes less chaotic,
$k_{\rm min}$ falls, more and more modes $a_k$ come into equilibrium, and the
energy scale ${\cal E}_{\rm equil}$ increases.  (iii) Maximum equilibration to
RMT is attained when $T_{\rm decay} \sim \hbar {\langle B^2 \rangle}^{-1/2}$,
where all modes $a_k$ for $k > k_{\rm min} \sim \hbar T_{\rm kick}^{-1}
{\langle B^2 \rangle}^{-1/2}$ equilibrate, and ${\cal E}_{\rm equil} \sim
{\langle B^2 \rangle}^{1/2}$.  (iv) Then, as the degree of chaoticity of the
system continues to decrease, $k_{\rm min}$ begins to increase, modes again
move away from RMT equilibrium and the energy ${\cal E}_{\rm equil}$ begins to
drop.  (v) Eventually, we reach the border $T_{\rm decay} \sim T_H$ between
quantum chaos and quantum regularity, where all equilibration to universal
statistics is again absent.  Thus, absence of equilibration to RMT at a given
time or energy scale may be consistent with either very strongly chaotic or
very weakly chaotic (or regular) systems, while maximum possible equilibration
is attained for moderately chaotic systems between these two extremes. 

\begin{figure}[ht]
\centerline{
\psfig{file=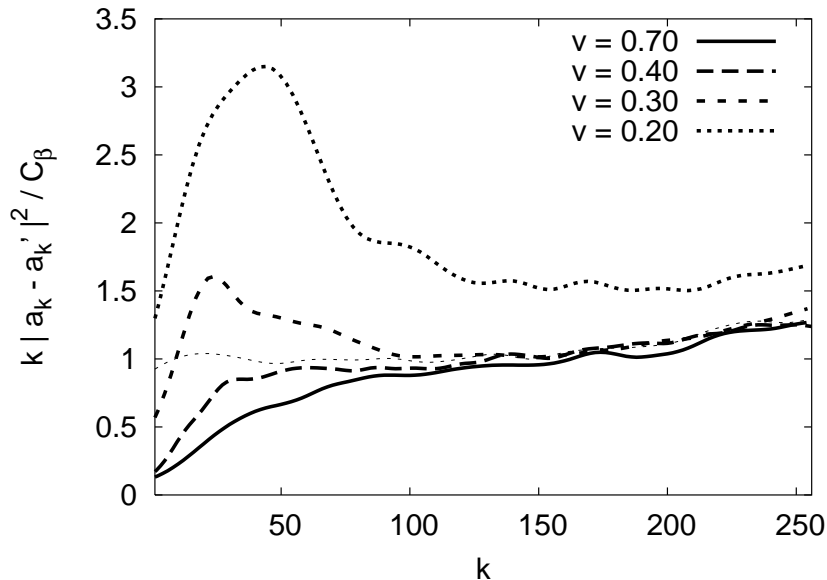,angle=270,width=4.5in}}
\vskip 0.2in
\caption{
The effect of a perturbation on the quantum spectrum as a function of Fourier
mode $k$ is plotted for four classical systems, distinguished by the parameter
$v$ of Eq.~(\ref{sawtooth}).  The values of $K_1$ and $K_2$ are as in
Fig.~\ref{fig_clas}, and the Hilbert space dimension is $N=512$.  As in the
previous figure, the data are averaged over an ensemble of perturbations
$B(q,p)=\sum_{m=1}^3 \left [ x_m \sin(2\pi m q)+ y_m \cos(2 \pi m q) \right ]$
with $\langle B^2 \rangle={1 \over 2} \sum_{m=1}^3 (x_m^2+y_m^2)=0.6
N^{-3}=4.51\cdot 10^{-9} $.  The constant $C_\beta$, is given by
Eq.~(\ref{akdiffrmt}), where $\beta=2$ in the present case.  The thin dotted
curve indicates the limit of full equilibration on all energy scales, as
obtained by choosing random and uncorrelated Hamiltonians $\hat H$ and $\hat
H'=\hat H+\hat B$.
}
\label{fig_a}
\end{figure}

 The discussion of the preceding paragraph is illustrated for the perturbed
sawtooth map in Fig.~\ref{fig_a}, where we show the effect on the spectrum at
various energy scales of a fixed-size perturbation, for four classical systems
characterized by different decay times $T_{\rm decay}$.  The classical systems
treated here are the same as in Fig.~\ref{fig_clas}, while the family of
perturbations used corresponds precisely to the family of classical functions
used in that earlier figure.  The squared change in each Fourier spectral
coefficient $a_k$ has been normalized by the constant $C_\beta$ of
Eq.~(\ref{akdiffrmt}), so that full equilibration implies $k |a_k - a_k'|^2
/C_\beta =1$ for $k \ll N$.  For $k \sim N$, this result is modified due to
higher-order terms in Eq.~(\ref{daexpand}), as discussed in
Section~\ref{secbrownian} and illustrated by the thin dotted curve in
Fig.~\ref{fig_a}.  From $v=0.70$ to $v=0.20$, the four systems clearly
demonstrate the effect of gradually reducing the Lyapunov exponent and
increasing the classical parameter $T_{\rm decay}$.

 Looking first at the $v=0.70$ curve in Fig.~\ref{fig_a}, we find equilibration
of the spectrum only for $k \ge 100$, corresponding approximately to energy
scales of three levels spacings or fewer. When the degree of chaos is reduced
($v=0.40$) and $T_{\rm decay}$ correspondingly increases, the same perturbation
strength is sufficient to equilibrate the spectrum up through much larger
energy scales, namely $k \ge k_{min} \approx 50$.  However, further reducing
the degree of chaos by tuning $v$ down below $0.4$ and thereby increasing
$T_{\rm decay}$ serves to {\it reduce} the range of equilibrated energies for
the same perturbation strength, as equilibration is now governed by the
condition $k \gg T_{\rm decay}/T_{\rm kick}$.  Finally, for $v=0.20$, $T_{\rm
decay}/T_{\rm kick} =120$, and once again equilibration to universal behavior
is not attained on any energy scale for our system size.

\section{Equilibration of Level Spacings} \label{secspac}

 In the previous section we considered spectral equilibration at various time
scales $k T_{\rm kick}$ and energy scales $E$.  We now focus specifically on
the conditions for equilibration at the energy scale $\Delta$, necessary for
example to reproduce the Wigner-Dyson distribution of nearest neighbor level
spacings.  Assuming $T_{\rm decay} < T_H$ as before, we need to satisfy
Eq.~(\ref{econd}) for ${\cal E} \sim \Delta$:
\begin{equation}
\langle B^2 \rangle \gg {\hbar \Delta \over T_{\rm decay}} \propto
{\hbar^{d+1} \over T_{\rm decay}} \,.
\end{equation}
The size of the perturbation $\hat B$ must generically be at least of order
$\hbar^{(d+1)/2}$ to produce equilibration of the level spacings.  Less chaotic
systems, however, as measured by a longer decay time, equilibrate more
efficiently.  For example, in a two-dimensional Hamiltonian system, $B \propto
\hbar^{3/2}$ is needed for a fixed $T_{\rm decay}$, but $B \propto \hbar^{7/4}$
would be sufficient if $T_{\rm decay} \propto T_H^{1/2}$, and $B \propto
\hbar^{2 - \epsilon /2}$ suffices if $T_{\rm decay} \propto T_H^{1-\epsilon}$.

\begin{figure}[ht]
\centerline{
\psfig{file=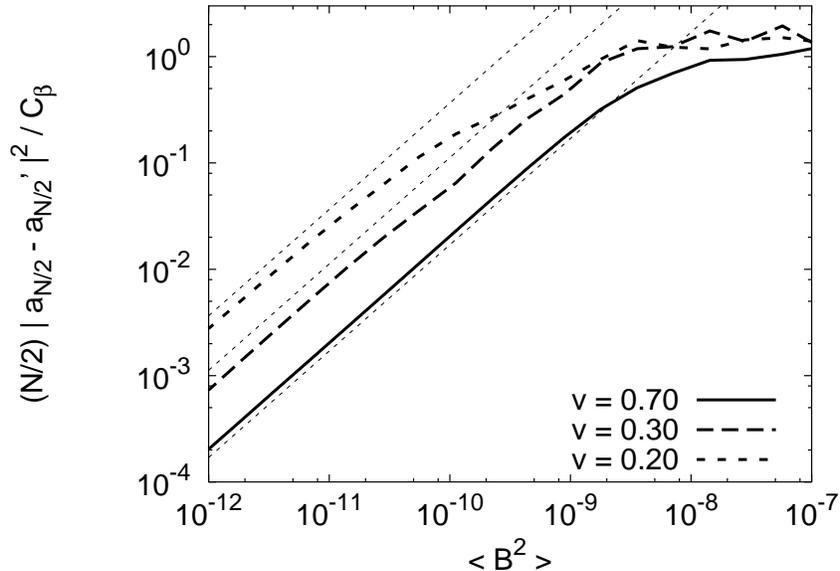,angle=270,width=4.5in}}
\vskip 0.2in
\caption{
Equilibration of the highest-frequency mode in the spectrum is studied for
$N=256$ as a function of perturbation strength for several initial
Hamiltonians.  The classical systems as well as the ensemble of perturbations
are the same as in Fig.~\ref{fig_a}.  The thin dotted lines correspond to the
classical prediction $|a_{N/2}-a'_{N/2}|  \propto \langle B^2 \rangle
\int_0^\infty dt \,P(t)$.
}
\label{fig_str}
\end{figure}

 Equilibration at the scale of the mean level spacing may be measured by
focusing on the Fourier coefficient $a_{N/2}$ (notice that $a_N=a_0$).
Equilibration of this coefficient is studied in Fig.~\ref{fig_str} as a
function of perturbation strength, for several classical Hamiltonians.
Specifically, the mean squared fluctuation in the $a_{N/2}$ coefficient is
plotted as a function of perturbation strength $\langle B^2 \rangle$.  For each
classical system, we clearly observe the proportionality between
$|a_{N/2}-a'_{N/2}|^2$ and $\langle B^2 \rangle$ when $|a_{N/2}-a'_{N/2}|^2 \ll
N^{-1}$, and the eventual saturation at the system-independent equilibrium
value at large perturbation size, just as predicted by the Brownian motion
model.  Furthermore, at a given perturbation strength, we see faster
equilibration for systems with slower classical correlation falloff, as
indicated by smaller classical parameter $v$.  Semiclassically, we expect this
increase in equilibration rate to be controlled by the integral $\int_0^\infty
dt \, P(t) \approx b T_{\rm decay}$ as in Eqs.~(\ref{sigmaint}) and
(\ref{longdom}); the classical predictions are indicated by dotted lines in
Fig.~\ref{fig_str}.  The slight discrepancy between the quantum data and the
semiclassical prefactors may be explained by the finite system size, $N=256$,
since the classical expressions assume $N \gg T_{\rm decay}/T_{\rm kick}$,
while in our case $N=256$ and $T_{\rm decay}/T_{\rm kick}$ reaches $120$ when
$v=0.2$.

\begin{figure}[ht]
\centerline{
\psfig{file=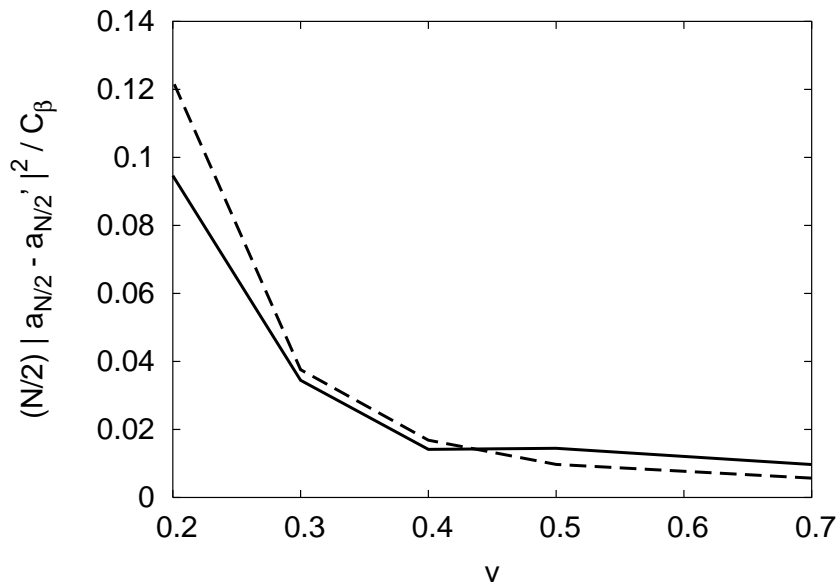,angle=270,width=4.5in}}
\vskip 0.2in
\caption{
Equilibration of the highest-frequency mode in the spectrum is studied, for
$N=1024$ and perturbation strength $\langle B^2 \rangle =6.4 \cdot 10^{-7}$, as
a function of the classical parameter $v$ (solid curve).  The classical systems
and the ensemble of perturbations are the same as in Figs.~\ref{fig_a} and
\ref{fig_str}.  The classical prediction based on Eq.~(\ref{sigmaint}) is
indicated by the dashed curve.
}
\label{fig_c}
\end{figure}

 Fig.~\ref{fig_c} shows explicitly the dependence of equilibration rate on the
classical system dynamics, by varying the classical parameter $v$ while the
perturbation strength is held fixed.  Again, we see an order of magnitude
change in the equilibration rate as $v$ is varied, in agreement with the
semiclassical prediction indicated by the dashed curve.

\section{Quantization Ambiguity and Spectral Equilibration}
\label{secambig}

 The original motivation behind Wilkinson's work~\cite{wilk} was to understand
the relationship between spectral equilibration and the quantization ambiguity,
thus using the latter to explain the approach of generic chaotic systems to RMT
behavior on short energy scales.  Generally, quantization ambiguities come in
two types.  The first are gauge or boundary condition ambiguities that allow
many different semiclassical theories to be associated with the same classical
dynamics, as in the Aharonov-Bohm effect or in a change from Dirichlet to
Neumann boundary conditions for a billiard system.  These ambiguities
correspond to $O(\hbar_{\rm eff})$ terms in the quantum Hamiltonian, where
$\hbar_{\rm eff}=\hbar/S_{\rm typ} \sim \hbar/ET_{\rm kick} \sim \hbar/pL$,
$S_{\rm typ}$ is the typical action of a short classical orbit, $p$ is the
typical momentum, and $L$ is the system size.  Secondly, there are operator
ordering ambiguities that allow multiple quantum Hamiltonians to have the same
semiclassical limit, including identical action phases at leading order.
Ambiguities of this second class are $O(\hbar_{\rm eff}^2)$ and result, for
example, from canonically quantizing the same classical Hamiltonian in two
coordinate systems related by a nonlinear canonical transformation~\cite{wilk};
they also appear naturally when different limiting procedures are used to
define a quantum dynamics on a constrained surface~\cite{ambig}.

 We first examine spectral equilibration due to the $O(\hbar_{\rm eff}^2)$
quantization ambiguity, considered by Wilkinson in Ref.~\cite{wilk}.
Substituting $B \sim \hbar_{\rm eff}^2 E$ into Eq.~(\ref{twoconde}), we find
equilibration on energy scales
\begin{equation}
\label{h2e}
{\cal E} \ll {\cal E}_{\rm equil}  = E \,{\rm min} \left ( \hbar_{\rm eff}
{T_{\rm kick} \over T_{\rm decay}}, \hbar_{\rm eff}^3 {T_{\rm decay} \over
T_{\rm kick}} \right )\,,
\end{equation}
or equivalently for Fourier modes
\begin{equation}
\label{h2k}
k \gg k_{\rm min} = {\rm max} \left ( {T_{\rm decay} \over T_{\rm kick}},
\hbar_{\rm eff}^{-2} {T_{\rm kick} \over T_{\rm decay}} \right ) \,.
\end{equation}
For a fixed classical dynamics, the second expression in the parentheses in
Eq.~(\ref{h2e}) or Eq.~(\ref{h2k}) always dominates in the semiclassical limit
$\hbar_{\rm eff} \to 0$.  For a $d$-dimensional Hamiltonian system, the total
number of available fluctuating modes in the spectrum is $O(\hbar_{\rm
eff}^{1-d})$, and all but the first $O(\hbar_{\rm eff}^{-2} T_{\rm kick}/T_{\rm
decay})$ of these are equilibrated.  Thus, full equilibration is not possible
for any spectral mode in the case $d=2$, while for $d=3$ full equilibration is
possible for the highest-$k$ modes, assuming slow decay of classical
correlations, $T_{\rm decay}/T_{\rm kick} \gg 1$.  In the energy domain, this
implies equilibration only on scales up to $O(T_{\rm decay}/T_{\rm kick})$ mean
level spacings $\Delta$ in the $3$-dimensional case.  The case $d=4$ (e.g., two
interacting particles in two dimensions) is the first for which equilibration
generically extends to scales much larger than a mean level spacing, ${\cal
E}_{\rm equil} \gg \Delta$.  Finally, in the many-body limit $d \to \infty$, an
ever-increasing number of modes are equilibrated by the $O(\hbar_{\rm eff}^2)$
ambiguity, however, the first $O(\hbar_{\rm eff}^{-2} T_{\rm kick}/ T_{\rm
decay})$ modes are never equilibrated.  Correspondingly the range of energies
over which equilibration may occur always remains a factor of $O(\hbar_{\rm
eff}^{2} T_{\rm decay}/ T_{\rm kick})$ smaller than the ballistic Thouless
energy $E_{\rm Thouless} \sim \hbar_{\rm eff}E$, independent of dimension.

 We now turn to the $O(\hbar_{\rm eff}^{-1})$ quantization ambiguity,
associated with external gauge fields or boundary conditions.  A similar
analysis shows that equilibration now occurs on energy scales ${\cal E} \ll
{\cal E}_{\rm equil} \sim \hbar_{\rm eff}E T_{\rm kick}/T_{\rm decay}$ and for
Fourier modes $k \gg k_{\rm min} \sim T_{\rm decay}/T_{\rm kick}$.  Thus, all
modes equilibrate in any dimension for a chaotic system, except for the first
few that encode non-universal short-time dynamics for a classical system with
slow correlation decay.

\begin{figure}[ht]
\centerline{
\psfig{file=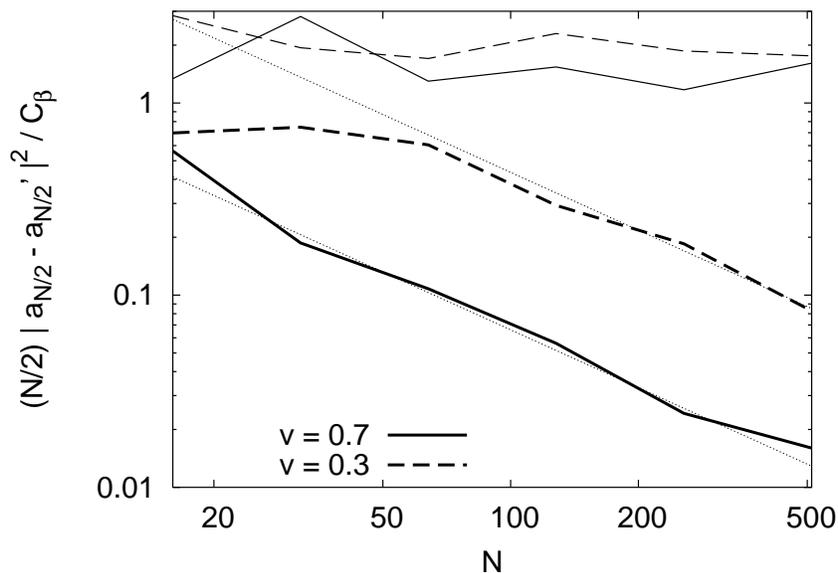,angle=270,width=4.5in}}
\vskip 0.2in
\caption{
Equilibration of the highest-frequency mode in the spectrum of a perturbed
sawtooth map is shown for $O(\hbar_{\rm eff}^2)$ perturbation strength $\langle
B^2 \rangle =(0.66/N^2)^2$ (thick curves) and $O(\hbar_{\rm eff})$ perturbation
strength $\langle B^2 \rangle =(0.16/N)^2$ (thin curves) as a function of $N$.
The thin dotted lines indicate the predicted $N^{-1}$ behavior for the
$O(\hbar_{\rm eff}^2)$ case in the semiclassical limit $N \to \infty$.
}
\label{fig_b}
\end{figure}

 In Fig.~\ref{fig_b}, we again focus on equilibration of the highest-frequency
Fourier mode $a_{N/2}$, corresponding to energy scales of order $\Delta$, for
our quantum map numerical model (which exhibits the scaling of a $d=2$
autonomous Hamiltonian system).  Consistently with the above discussion, an
$O(\hbar_{\rm eff}^2)$ ambiguity is not sufficient to equilibrate even this
highest-frequency Fourier mode in the semiclassical limit $N \to \infty$
($\hbar_{\rm eff} \to 0$); lower-frequency modes will be even further away from
equilibration.  On the other hand, an $O(\hbar_{\rm eff})$ (gauge-size)
ambiguity provides full equilibration independent of $\hbar_{\rm eff}$.

 The above analysis and numerical simulation assume a constant $T_{\rm decay}$
in the $\hbar_{\rm eff} \to 0$ limit.  However, a system remains in the quantum
chaotic regime as long as $T_{\rm decay}$ is shorter than the Heisenberg time,
i.e. when $T_{\rm decay}/T_{\rm kick} < \hbar_{\rm eff}^{1-d}$.  In general,
the rate of equilibration to RMT behavior is increased by choosing a system
with very slow classical relaxation, $T_{\rm decay} \gg T_{\rm kick}$.  For the
$d=2$ case, it is impossible to attain complete equilibration to RMT for any
spectral mode with an $O(\hbar_{\rm eff}^2)$ perturbation, regardless of the
choice of classical system.  For $d\ge 3$, on the other hand, Eqs.~(\ref{h2e})
and (\ref{h2k}) clearly indicate that full equilibration for some modes can be
achieved with an $O(\hbar_{\rm eff}^2)$ perturbation for systems with $T_{\rm
decay}/T_{\rm kick} \gg 1$ (where a generic chaotic system with $T_{\rm
decay}/T_{\rm kick} \sim 1$ would not display equilibrated behavior).  Optimal
equilibration is attained for $T_{\rm decay}/ T_{\rm kick} \sim \hbar_{\rm
eff}^{-1}$, where all but the lowest $O(\hbar_{\rm eff}^{-1})$ modes are
equilibrated in any dimension $d \ge 3$.

\begin{figure}[ht]
\centerline{
\psfig{file=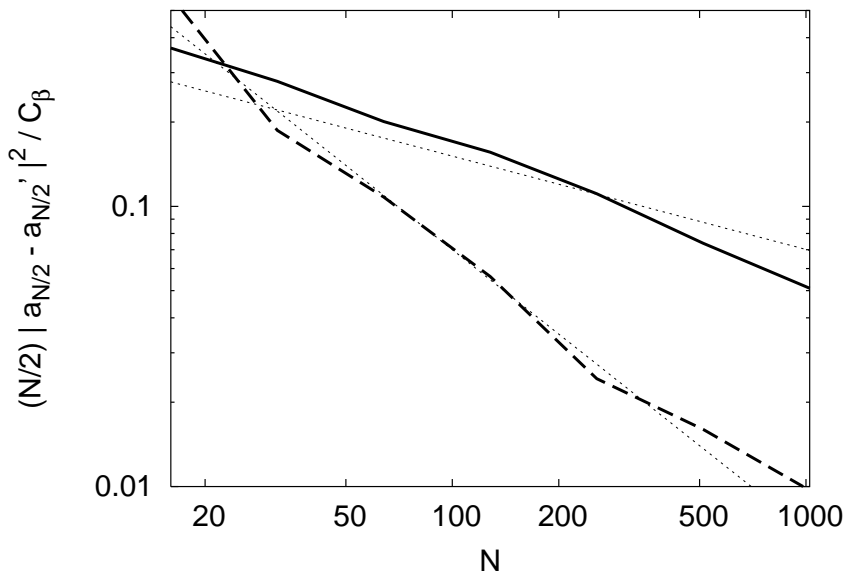,angle=270,width=4.5in}}
\vskip 0.2in
\caption{
Equilibration of the highest-frequency mode in the spectrum is shown for
$O(\hbar^2)$ perturbation strength $\langle B^2 \rangle =(0.66/N^2)^2$, for a
system with power-law classical correlation decay, defined by
Eq.~(\ref{clpower}), solid curve, and for the perturbed sawtooth map with
$v=0.7$, dashed curve.  The thin dotted lines indicate the predicted
$N^{-\gamma}$ behavior, for $\gamma=1/3$ and $\gamma=1$.
}
\label{fig_bb}
\end{figure}

 Finally, in Fig.~\ref{fig_bb} we compare equilibration in the hard chaotic
perturbed sawtooth map with equilibration in a system governed by a slow
power-law classical decay, defined by Eq.~(\ref{clpower}).  In both cases a
perturbation of order $\hbar_{\rm eff}^2 \sim N^{-2}$ is used.  Although no
equilibration is possible for any system in $d=2$, we see clearly that the
power-law classical system comes closer to equilibration than the hard chaotic
system for any large $N$, and the two systems are governed by different
exponents in the $N \to \infty$ limit.

\section{Fidelity and Semiclassical Accuracy}
\label{secsemi}

 The fidelity of quantum evolution in the presence of a small perturbation,
also known as the Loschmidt echo, has received much attention recently,
particularly in the context of classical--quantum correspondence and the very
different behaviors that the quantum fidelity can exhibit for classically
regular and chaotic systems~\cite{emerson}.  For sufficiently large
perturbations, the decay of quantum fidelity with time can be governed by the
classical Lyapunov exponent~\cite{jalabert,jacquod}.  Here we wish to mention
an interesting connection between spectral equilibration and fidelity decay in
quantum chaotic systems due to a {\it small} perturbation, as analyzed recently
by Prosen and collaborators~\cite{prosen}.  It was shown that for a static
perturbation $\hat B$ with $\langle B^2 \rangle < \hbar^2/T_{\rm decay}^2$, the
fidelity decays exponentially on a time scale varying inversely with $T_{\rm
decay}$, namely $T_{\rm fidelity} \sim \hbar^2/\langle B^2 \rangle T_{\rm
decay}$.  Dividing through by the one-step time scale $T_{\rm kick}$, we find
that $T_{\rm fidelity}$ corresponds to a dimensionless mode number
\begin{equation}
k_{\rm fidelity} \sim {\hbar^2 \over \langle B^2 \rangle T_{\rm decay} T_{\rm
kick}} \,, \end{equation}
which precisely agrees with the boundary between equilibrated and
non-equilibrated modes given by Eq.~(\ref{kcond}), valid as long as $k_{\rm
fidelity} > T_{\rm decay}/T_{\rm kick}$.  In other words, a mode $k$ is
equilibrated to universal behavior by a given class of perturbations if and
only if two conditions hold simultaneously: the quantum fidelity has decayed to
a value much less than unity by the corresponding time $k T_{\rm kick}$ {\it
and} this time scale $k T_{\rm kick}$ is larger than the classical relaxation
time $T_{\rm decay}$. 

 Another important connection is between spectral equilibration behavior
discussed in the present work and the decay of semiclassical accuracy.  Since
different quantizations of the same semiclassical dynamics differ by
$O(\hbar_{\rm eff}^2)$ in the Hamiltonian, the difference between a typical
quantization and the semiclassical approximation must be at least of this order
in the $\hbar_{\rm eff} \to 0$ limit.  Thus the error in the semiclassical
approximation must be at least of the same size as the error caused by an
$O(\hbar_{\rm eff}^2)$ perturbation in quantum mechanics, assuming of course
that the physically correct quantization is ``typical".  Clearly, hard quantum
effects such as diffraction can increase the error in the semiclassical
approximation, so the $O(\hbar_{\rm eff}^2)$ quantization ambiguity only
provides a lower bound on the size of the semiclassical error, or equivalently
an upper bound on the breakdown time of semiclassical validity.  Previous work
has shown that this bound on the semiclassical error is saturated for the case
of a smooth Hamiltonian in the absence of caustics~\cite{sckaplan}, and
furthermore it was demonstrated that the semiclassical error obeys different
scaling laws with time and $\hbar$ for regular as opposed to chaotic classical
systems, just as one would predict using the quantization ambiguity approach.
The semiclassical accuracy problem may also be considered as an example of a
generalized quantum fidelity problem, if non-Hermitian perturbations are
considered (since semiclassical evolution is in general non-unitary).

 The results of the present work, particularly those of Sec.~\ref{secambig},
imply an upper bound on the breakdown time scale of semiclassical accuracy:
$T_{\rm semiclassical} \sim \hbar_{\rm eff}^{-2} T_{\rm kick}^2/T_{\rm decay}$
and a lower bound on the breakdown energy scale of semiclassical accuracy:
\begin{equation}
{\cal E}_{\rm semiclassical} \sim E \hbar_{\rm eff}^3 {T_{\rm decay} \over
T_{\rm kick}} \sim E_{\rm Thouless} \hbar_{\rm eff}^2 {T_{\rm decay} \over
T_{\rm kick}} \,,
\end{equation}
where $E_{\rm Thouless}$ is a ballistic Thouless energy.  In particular,
semiclassical accuracy will persist beyond the Heisenberg time $T_H \sim
\hbar_{\rm eff}^{-1} T_{\rm kick}$ in any two-dimensional chaotic system
(assuming diffraction and caustics are properly accounted for), allowing
individual energy levels and wave functions to be semiclassically resolved.  In
three-dimensional systems, $T_H \sim \hbar_{\rm eff}^{-2} T_{\rm kick}$, and
$T_{\rm semiclassical} \sim T_H T_{\rm kick}/T_{\rm decay}$.  Here, the
semiclassical breakdown time may be comparable to the Heisenberg time for the
most chaotic systems ($T_{\rm decay}/T_{\rm kick} \sim 1$), but any slowdown in
the classical relaxation rate will cause a faster breakdown in the
semiclassical approximation.  Finally, given four or more classical degrees of
freedom, semiclassical accuracy inevitably breaks down well before the
Heisenberg time for any generic dynamics, even the most chaotic, and
semiclassical reproduction of individual spectral levels is never possible.  It
may be of interest to investigate the manner in which improved semiclassical
approximations beyond leading order in $\hbar$~\cite{weibert} may produce
different scaling of the semiclassical error and possibly permit quantum
spectra to be semiclassically resolved in four dimensions and higher.

\section{Summary}
\label{secsummary}

 A careful examination of the Brownian motion model for spectral equilibration
to universal statistics shows that equilibration is strongly dependent on the
classical relaxation rate as well as on the dimensionality of the system.  Two
competing time scales, proportional and inversely proportional to the classical
relaxation time $T_{\rm decay}$, jointly govern the equilibration process.
Balancing of these two time scales implies that for a given perturbation $B$,
equilibration of the maximum number of spectral modes is achieved when $T_{\rm
decay} \sim \hbar \langle B^2 \rangle^{-1/2}$.  For small perturbations, $B <
\hbar/T_{\rm decay}$, a relation exists between spectral equilibration as a
function of mode number $k$ and the decay of the quantum fidelity as a function
of time.

 Focusing on the effect of $O(\hbar_{\rm eff}^2)$ perturbations, associated
with the ambiguity of quantization, we find that no equilibration to universal
statistics is ever possible for dynamics in two dimensions.  In three
dimensions, equilibration to universal statistics occurs for some modes, but
only if the classical relaxation time $T_{\rm decay}$ is sufficiently long,
while in four dimensions and higher, equilibration of at least some modes
occurs for any chaotic system.  However, optimal equilibration only occurs for
very long classical relaxation times, $T_{\rm decay} \propto \hbar_{\rm
eff}^{-1}$.  These predictions of the Brownian motion model are also entirely
consistent with results for semiclassical accuracy in smooth chaotic systems.

\section*{Acknowledgments} The author is grateful to B.~Mehlig for invaluable
contributions in the early stages of this work, and to M.~Wilkinson and
U.~Smilansky for very useful discussions.  This work was supported in part by
the U.S. Department of Energy Grant No.\ DE-FG03-00ER41132 and the Louisiana
Board of Regents Support Fund Contract LEQSF(2004-07)-RD-A-29.

\end{document}